\documentclass[aps,superscriptaddress,amsmath,amssymb,floatfix,twocolumn,showpacs,amsfonts,longbibliography]{revtex4-2}
\usepackage{times}
\usepackage[varg]{txfonts}
\usepackage{textcomp}
\usepackage{graphicx}
\usepackage{subfigure}
\usepackage{color}
\usepackage[colorlinks=true,citecolor=blue,urlcolor=blue,linkcolor=blue,hyperindex]{hyperref}
\usepackage{braket}
\usepackage{overpic}
\usepackage{cancel}
\usepackage{mhchem}
\usepackage[normalem]{ulem}

\def\bp{\mathbf{p}}
\def\bk{\mathbf{k}}

\def\bS{\mathbf{S}}

\def\bB{\mathbf{B}}
\def\bb{\mathbf{b}}
\def\btb{\mathbf{\tilde{b}}}
\def\tb{\tilde{b}}
\def\bgk{\boldsymbol{\gamma}_\mathbf{k}}
\def\bdk{\mathbf{d}_{\mathbf{k}}}
\def\bd{\mathbf{d}}
\def\bR{\mathbf{R}}

\def\up{\uparrow}
\def\down{\downarrow}
\def\non{\nonumber}
\def\half{\frac{1}{2}}

\def\e{\varepsilon}
\def\TK{T_K}
\def\Tc{T_c}
\def\Tco{T_{c0}}

\def\iwn{i\omega_n}

\definecolor{gr}{RGB}{0,187,0}

\begin{document}

\title{A unified theoretical framework for Kondo superconductors: Periodic Anderson impurities with attractive pairing and Rashba spin-orbit coupling}
\author{Shangjian Jin}
\affiliation{Department of Physics, National University of Singapore, 2 Science Drive 3, 117551 Singapore}
\affiliation{Centre for Advanced 2D Materials, National University of Singapore, 6 Science Drive 2, 117546 Singapore}

\author{Darryl C. W. Foo}
\affiliation{Centre for Advanced 2D Materials, National University of Singapore, 6 Science Drive 2, 117546 Singapore}

\author{Tingyu Qu}
\affiliation{Department of Physics, National University of Singapore, 2 Science Drive 3, 117551 Singapore}

\author{Barbaros \"{O}zyilmaz}
\affiliation{Department of Physics, National University of Singapore, 2 Science Drive 3, 117551 Singapore}
\affiliation{Centre for Advanced 2D Materials, National University of Singapore, 6 Science Drive 2, 117546 Singapore}
\affiliation{Department of Materials Science and Engineering, 
National University of Singapore, 9 Engineering Drive 1, 
Singapore 117575}

\author{Shaffique Adam}
%\email[Corresponding author:]{shaffique@wustl.edu}
\affiliation{Department of Materials Science and Engineering, 
National University of Singapore, 9 Engineering Drive 1, 
Singapore 117575}
\affiliation{Department of Physics, Washington University in St. Louis, St. Louis, Missouri 63130, United States}

\date{\today}

\begin{abstract}
Magnetic superconductors manifest a fascinating interplay between their magnetic and superconducting properties. This becomes evident, for example, in the significant enhancement of the upper critical field observed in uranium-based superconductors, or the destruction of superconductivity well below the superconducting transition temperature $\Tc$ in cobalt-doped NbSe$_2$. In this work, we argue that the Kondo interaction plays a pivotal role in governing these behaviors. By employing a periodic Anderson model, we study the Kondo effect in superconductors with either singlet or triplet pairing.  In the regime of small impurity energies and high doping concentrations, we find the emergence of a Kondo resistive region below $\Tc$.  While a magnetic field suppresses singlet superconductivity, it stabilizes triplet pairing through the screening of magnetic impurities, inducing reentrant superconductivity at high fields.  Moreover, introducing an antisymmetric spin-orbital coupling suppresses triplet superconductivity.  This framework provides a unified picture to understand the observation of Kondo effect in NbSe$_2$ as well as the phase diagrams in Kondo superconductors such as UTe$_2$, and URhGe.
\begin{description}
\item[PACS number(s)] 74.20.Fg, 74.20.Rp, 74.25.Dw
\end{description}
\end{abstract}

\maketitle

\section{Introduction}\label{sec:intro}

The intricate interplay between magnetism and superconductivity has garnered significant attention due to the competition of magnetic and superconducting ground states. Various lanthanum-based heavy fermion singlet superconductors~\cite{riblet1971vanishing,maple1972re,huber1974superconducting,winzer1977evidence} have shown a second transition temperature $\TK$ where a resistive state resurfaces due to the Kondo effect that disrupts superconductivity \cite{muller1971kondo,gorkov2005kondo}.  More recently, there has been a surge of interest in uranium-based ferromagnetic heavy fermion superconductors such as URhGe, UCoGe, and UTe$_2$ for their potential to realize spin triplet superconductivity with nontrivial topological properties~\cite{sato2017topoSCreview,aoki2019reviewUbased,ishizuka2019TopoSCUTe2,Shishidou2021topoute2,aishwarya2023magneticUTe2}. These compounds hold promise as candidates for a Kondo-effect-induced topological superconductors~\cite{ran2019nearly_UTe2,jiao2020chiralkondoUTe2,aoki2022unconventional,chang2024topokondoSC}. Moreover, the hybridization between itinerant electrons and localized 5$f$ electrons of uranium atoms allows for a coexistence of the Kondo effect and superconductivity that was predicted theoretically~\cite{suzuki2019groundkondo,suzuki2020kondo,kang2022U5fTe5p,thomas2023metamagnetic}, and confirmed experimentally~\cite{troc2010bulk,jiao2020chiralkondoUTe2,butch2022INSkondoute2}.  This underscores the importance of the Kondo interaction in elucidating the $B-T$ phase diagram of U-based superconductors. The phase diagrams of these materials exhibit diverse features: URhGe displays high-field reentrant superconductivity with the magnetic field $\bB\parallel b$-axis~\cite{huxley2005URhGe}; UCoGe shows an S-shaped upper critical field~\cite{Jacques2009UCoGe}; and UTe$_2$ exhibits a large upper critical field~\cite{ran2019nearly_UTe2}, significant field anisotropy, field reentrant superconductivity~\cite{jacques2019reentrantUTe2}, and multiple SC phases under pressure~\cite{jacques2020phaseUTe2}.  

In a separate development, few layer transition metal dichalcogenides have recently emerged as a feasible and customizable platform conducive to two-dimensional Kondo superconductivity, explored both experimentally~\cite{devidas2023QDNbSe2,Ugeda2023kondoTMD} and theoretically~\cite{Vivek2017tmd,sticlet2019FMchain_NbSe2,zhang2020kondo}.  For example, in a recent experiment~\cite{Tingyu2023CoNbSe2}, NbSe$_2$ encapsulated by atomically-thin boron nitride was interfaced with magnetic dopants introduced by e-beam evaporation of Co.  The observed destruction of superconductivity concomitant with a logarithmic resistance-temperature behavior was attributed to the Kondo effect.  In this work, we refer to this platform as Co-NbSe$_2$.  We note that the superconducting state of few-layers NbSe$_2$ is intrinsically parity mixed due to Ising spin-orbit coupling~\cite{gor_rashba2001superconducting,mockli2018NbSe2,he2018nodalNbSe2,mazin2020FMinstability,pribiag2021fewlayernbse2,wan2023orbitalFFLO}. It is believed that manipulating the Kondo effect in NbSe$_2$ through doping concentration, temperature, and magnetic field could pave the way for constructing pure triplet superconducting spintronic devices (see e.g. Ref.~\cite{ara2023control}). 

Despite the similarity in the underlying physics of these different systems, to our knowledge there is no unified microscopic theory to explain the diverse $B-T$ phase diagrams observed in materials ranging from Co-NbSe$2$ to U-based superconductors.  This is the goal of the present work.  Before we outline our model, we briefly discuss what is known in the theoretical literature.  More than fifty years ago, M{\"u}ller-Hartmann and Zittartz~\cite{muller1971kondo} predicted the destruction of superconductivity by the Kondo effect using the Nagaoka approach.  This mechanism fails below the Kondo temperature and always predicts superconductivity independent of the impurity concentration.  About 20 years ago, Barzykibn and Gor'kov~\cite{gorkov2005kondo} studied the Kondo effect in s-wave superconductors using a periodic Anderson model elucidating the relationship between $\Tc$ and the impurity concentration for Ce$_{1-x}$La$_x$Ru$_3$Si$_2$, but they did not consider the effect of magnetic fields.  Almost five years ago, Suzuki and Hattori~\cite{suzuki2020kondo} explored a possible connection between Kondo coupling and reentrant triplet superconductivity using a one-dimensional Kondo lattice model.  More recent work by Machida~\cite{machida2021nonunitary} established a Ginzburg-Landau theory considering the rotation of the triplet order parameter $\bdk$ vector to qualitatively determine the $B_{c2}$ shape in URhGe, UCoGe and UTe$_2$, while assuming a temperature-dependent absolute upper field limit.  It is in this context that we develop a comprehensive theory capable of capturing the roles of magnetic impurities and magnetic fields in both singlet and triplet superconductors to understand the diverse patterns of superconductivity destruction and reentrance across different materials.

To this end, we start with a periodic Anderson model to investigate the pair-breaking process in superconductors. Utilizing the Bardeen–Cooper–Schrieffer (BCS) weak coupling assumption and Green’s function approach, we solve the gap equations and derive pair-breaking equations.  For both singlet and triplet superconductors, this formalism predicts a second transition $\TK$ back to a normal state for a range of doping concentrations.  We are then able to incorporate both an external magnetic field and spin-orbit coupling.  This allows us to extract the phase-diagrams for a range of materials and predict some unexpected phenomena such as reentrant superconductivity for singlet superconductors, and the enhancement of $\Tc$ with magnetic fields for triplet superconductors.

%%%%%%%%%%%%%%%%%%%%%%%%%%%%
\begin{figure*}[t]
\centering\includegraphics[width=7.0in]{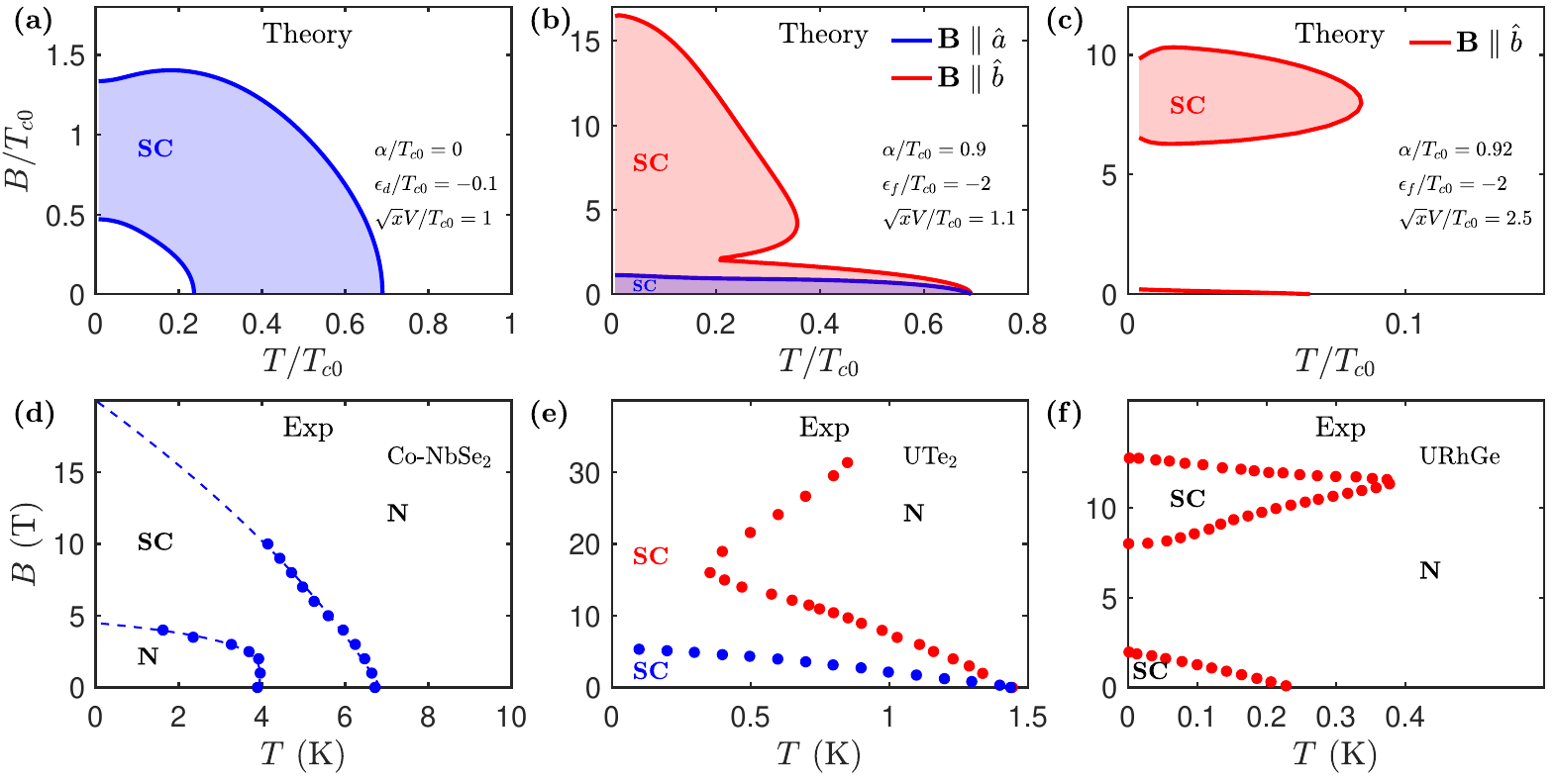}
\caption{Theoretical calculation for the superconducting $B-T$ phase diagrams (upper panels) compared with experimental results (bottom panels) for Co-NbSe$_2$, UTe$_2$, and URhGe.  Qualitative agreement with experiment is achieved within the same theoretical framework. (a) Singlet superconducting boundary that is isotropic in magnetic field. (b) The blue (red) line is the computed triplet superconducting boundary with $\bB$ parallel to $a$-axis ($b$-axis) using $\bgk=\alpha(-k_y,k_x,0)$ and $\bdk=(k_y+i k_z,k_x,i k_x)$. (c) Triplet superconducting boundary using $\bgk=\alpha(0,k_x,0)$ and $\bdk=(k_z,ik_z,k_x+ik_y)$. (d) The experimental data of Co-NbSe$_2$ for $\bB\parallel$ $ab$ plane is from Ref.~\cite{Tingyu2023CoNbSe2}. (e) and (f) Data for UTe$_2$ and URhGe are extracted from Ref.~\cite{aoki2022unconventional} and Ref.~\cite{huxley2005URhGe}, respectively.
}
\label{fig:theo_exp}
\end{figure*}
%%%%%%%%%%%%%%%%%%%%%%%%%%%%%%%%%%%%%%

\section{Periodic Anderson model}\label{sec:PAM}

We start with the periodic Anderson model, a standard approach for describing many transition metal, rare-earth, and heavy fermion systems \cite{gschneidner2011PAM,held2019PAM_QCP,chang2024topokondoSC,Miyao2023PAM}
\begin{equation}\label{Eq:H0}
H_0=\sum_{\bk,s}{\e_\bk c_{\bk s}^\dag c_{\bk s}}+\sum_{i,s}{\e_f f_{is}^\dag f_{is}}+\frac{U}{2}\sum_{i}{f_{i\up}^\dag f_{i\down}^\dag f_{i\down} f_{i\up}}.
\end{equation}
\noindent Here $c_{\bk s}^\dag$ ($c_{\bk s}$) and $f_{i s}^\dag$ ($f_{i s}$) are the creation (annihilation) operators for the conduction electrons and the localized $f$-electrons, respectively. $\e_\bk$ is the conduction band dispersion and $\e_f$ denotes the impurity energy level. Both $\e_\bk$ and $\e_f$ are defined relative to the Fermi level. Since the repulsive on-site Coulomb potential is very strong for the $f$-electrons, we take it to be infinite in this work.  This is the Kondo limit where the doubly occupied $f$-electron states are projected out and the effective energy $\e_f<0$~\cite{Kondolimit_rice1985gutzwiller,chang2024topokondoSC}.  The hybridization between the local and conduction electrons can be written as 
\begin{equation}\label{Eq:Hcf}
H_{cf}=\sum_{i,\bk,s}{\Big( \sqrt{x}V e^{i\bk\cdot\bR_i} f_{i s}^\dag c_{\bk s}+h.c. \Big)},
\end{equation}
where the hybridization $V$ is assumed to be $\bk$-independent for simplicity.  Importantly, we have introduced $x=\braket{f_i^\dag f_i}$ that allows us to change the $f$-electrons concentration.  This corresponds to a virtual crystal approximation where each unit cell is considered to contain $f$-electrons with concentration $x$, resulting in an effective hybridization of $\sqrt{x} V$~\cite{gorkov2005kondo}. When the impurity energy is comparable to $k_B\Tco$ 
(which is much smaller than the bandwidth, on the order of 1 eV), the virtual crystal approximation is identical with the coherent potential approximation~\cite{Ehrenreich1968VCACPA,chen1973CPA} that replaces the inhomogeneous potential of a disordered material with an effective potential.

To study Kondo superconductors, we add an effective attractive superconducting pairing potential in the weak-coupling limit~\cite{sigrist1991phenomenological}
\begin{equation}\label{Eq:H_pair}
H_{pair}=\half \sum_{\bk,\bk^\prime,s_1,s_2}{V_{\bk\bk^\prime} c_{-\bk s_2}^\dag c_{\bk s_1}^\dag c_{\bk^\prime s_1} c_{-\bk^\prime s_2}}.
\end{equation}

\noindent In this work, we assume a generic attractive potential $V_{\bp\bk}=-\sum_{l=0}^{\infty}4\pi V_l\sum_m{Y_{lm}(\Omega_\bp)Y_{lm}^*(\Omega_\bk)}$, where $V_l=V_l(k_F,k_F)$ is constant and $Y_{lm}$ are spherical harmonics \cite{anderson1961generalized,balian1963superconductivity}. Such effective potential gives rise to both spin singlet and triplet pairing, with $l=0,1,2,3$ for $s$-, $p$-, $d$-, and $f$-wave superconducting order parameters. Our full Hamiltonian has the form 
\begin{equation}\label{Eq:Htot}
H=H_0+H_{\mathrm cf}+H_{\mathrm{ASOC}}+H_{Z}+H_{\mathrm pair},
\end{equation}
where $H_{\mathrm{ASOC}}$ allows for an antisymmetric spin-orbit coupling (ASOC)
in these compounds caused by the breaking of inversion symmetry,
\begin{equation}\label{Eq:H_ASOC}
H_{\mathrm {ASOC}}=\sum_{\bk,s_1,s_2}{\bgk\cdot\boldsymbol{\sigma}_{s_1 s_2}c_{\bk s_1}^\dag c_{\bk s_2}}
\end{equation}
which is of the Rashba type~\cite{gor_rashba2001superconducting,sigrist2004superconductivity}, and the magnetic field is included through a Zeeman field 
\begin{equation}\label{Eq:H_Zeeman}
H_{Z}=\sum_{\bk,s_1,s_2}{\bB \cdot\boldsymbol{\sigma}_{s_1 s_2}\bigg(c_{\bk s_1}^\dag c_{\bk s_2}+f_{\bk s_1}^\dag f_{\bk s_2}\bigg)},
\end{equation}
where $\bB$ is external magnetic field, and $\boldsymbol{\sigma}$ are the Pauli matrices. For simplicity, we set the Land\'e factor $g=2$ and use dimensionless variables where $\mu_B=k_B=\hbar=1$. 

\section{Results}\label{sec:res}

To determine the superconducting phase diagram, we calculate the superconducting transition temperature $\Tc$ by solving the gap equation. Following a Green's function approach for superconductivity, the linearized gap equation is given by~\cite{sigrist1991phenomenological}
\begin{equation}\label{Eq:Gapeq}
\Delta(\bp)=-T\sum_{n,\bk}{V_{\bp\bk}G_{cc}(\bk,\iwn)\Delta(\bk)G_{cc}^T(-\bk,-\iwn)},
\end{equation}
with the gap function $\Delta(\bk)=[\psi_\bk+\bdk\cdot \boldsymbol{\sigma}]i\sigma_y$, where $\psi_\bk$ ($\bdk$) is the singlet (triplet) order parameter. $G_{cc}(\bk,\iwn)$ is the normal state Green's function for conduction electrons, whose expression is shown in Appendix~\ref{sec:appendix_Green}. For simplicity we assume a spherical Fermi surface in all of our calculations. With this assumption, the mixed singlet and triplet gap equations become decoupled and can be solved separately as shown in Appendix~\ref{sec:appendix_gapeq}. The pair-breaking equations determining $\Tc$ are then given by 
% \begin{widetext}
% \begin{eqnarray}
% \label{Eq:main}
% \left\{\begin{array}{ll}
% \ln{\frac{\Tc}{T_{c0}}}=2\pi \Tc\sum_{n=0}^{\infty}{ \bigg\langle |\hat{\psi}_\bk|^2 \frac{\Omega_n \big[\Omega_n^2+\quarter(\bb-\btb)^2\big]}{\big[\Omega_n^2+\quarter(b+\tb)^2\big]\big[\Omega_n^2+\quarter(b-\tb)^2\big]} \bigg\rangle_\bk}-\sum_{n=0}^{\infty} \frac{1}{n+\half} & \mathrm{for \ singlet,}  \\ 
% \ln{\frac{\Tc}{T_{c0}}}=2\pi \Tc\sum_{n=0}^{\infty} \bigg\langle 
% \frac{|\hat{d}_\bk|^2\Omega_n \big[\Omega_n^2+\quarter(\bb+\btb)^2\big]-(\btb\cdot \hat{d}_\bk)(\bb\cdot \hat{d}_\bk^*)\Omega_n-i(\hat{d}_\bk \times \hat{d}_\bk^*) \cdot i[\half\Omega_n^2(\bb-\btb)+\frac{b^2-\tb^2}{8}(\bb+\btb)]}{\big[\Omega_n^2+\quarter(b+\tb)^2\big]\big[\Omega_n^2+\quarter(b-\tb)^2\big]}
% \bigg\rangle_\bk -\sum_{n=0}^{\infty} \frac{1}{n+\half} 
% & \mathrm{for \ triplet,} 
% \end{array}
% \right.
% \end{eqnarray}
% \end{widetext}

\begin{widetext}
\begin{eqnarray}
\label{Eq:main}
\left\{
\begin{array}{ll}
\displaystyle \ln{\frac{\Tc}{T_{c0}}}=2\pi \Tc\sum_{n=0}^{\infty} \bigg\langle |\hat{\psi}_\bk|^2 \frac{\Omega_n \big[\Omega_n^2+\frac{1}{4}(\bb-\btb)^2\big]}{\big[\Omega_n^2+\frac{1}{4}(b+\tb)^2\big]\big[\Omega_n^2+\frac{1}{4}(b-\tb)^2\big]} \bigg\rangle_\bk - \frac{1}{\omega_n}
& \mathrm{for \ singlet,}  \\[10pt]
\displaystyle \ln{\frac{\Tc}{T_{c0}}}=2\pi \Tc\sum_{n=0}^{\infty} \bigg\langle 
\frac{|\hat{d}_\bk|^2\Omega_n \big[\Omega_n^2+\frac{1}{4}(\bb+\btb)^2\big]-(\btb\cdot \hat{d}_\bk)(\bb\cdot \hat{d}_\bk^*)\Omega_n-i(\hat{d}_\bk \times \hat{d}_\bk^*) \cdot i\big[\half\Omega_n^2(\bb-\btb)+\frac{b^2-\tb^2}{8}(\bb+\btb)\big]}{\big[\Omega_n^2+\frac{1}{4}(b+\tb)^2\big]\big[\Omega_n^2+\frac{1}{4}(b-\tb)^2\big]}
\bigg\rangle_\bk - \frac{1}{\omega_n} 
& \mathrm{for \ triplet,} 
\end{array}
\right.
\end{eqnarray}
\end{widetext}

\noindent where we have introduced 
\begin{equation}\label{Eq:A}
\Omega_n=\omega_n+xV^2\mathrm{Im}\frac{\e_f-\iwn}{(\e_f-\iwn)^2-B^2}
\end{equation}
as the Matsubara frequencies shifted by doping, and the effective field 
\begin{equation}\label{Eq:b}
    \bb=\bgk+\bB[1+\frac{xV^2}{(\e_f-\iwn)^2-B^2}].
\end{equation}
\noindent Note that $b=\sqrt{\bb\cdot\bb}$ is a complex number whose argument is limited to $(-\frac{\pi}{2},\frac{\pi}{2}]$, and we denote $\mathbf{\tb}=\mathbf{b}(-\bk,-\iwn)$.  The normalized gap functions are $\hat{\psi}_\bk=\frac{\psi_\bk}{\sqrt{\braket{|\psi_\bk|^2}_\bk}}$ and $\hat{d}_\bk=\frac{\bdk}{\sqrt{\braket{|\bdk|^2}_\bk}}$, where the notation $\braket{\cdots}_\bk$ indicates an average over the Fermi surface. The bare superconducting critical temperature $\Tco^l=1.13\omega_D \exp[-\frac{1}{N(0)V_l}]$ for each channel is determined by the strength of effective attractive potential.  Finally, we have checked that Eq.~\eqref{Eq:main} reproduces the known results of Sigrist~\emph{et. al.}~\cite{sigrist2004superconductivity} in the limit of vanishing Kondo impurities i.e. $x=0$.

%%%%%%%%%%%%%%%%%%%%%%%%%%%%
\begin{figure}[t]
\centering\includegraphics[width=3.4in]{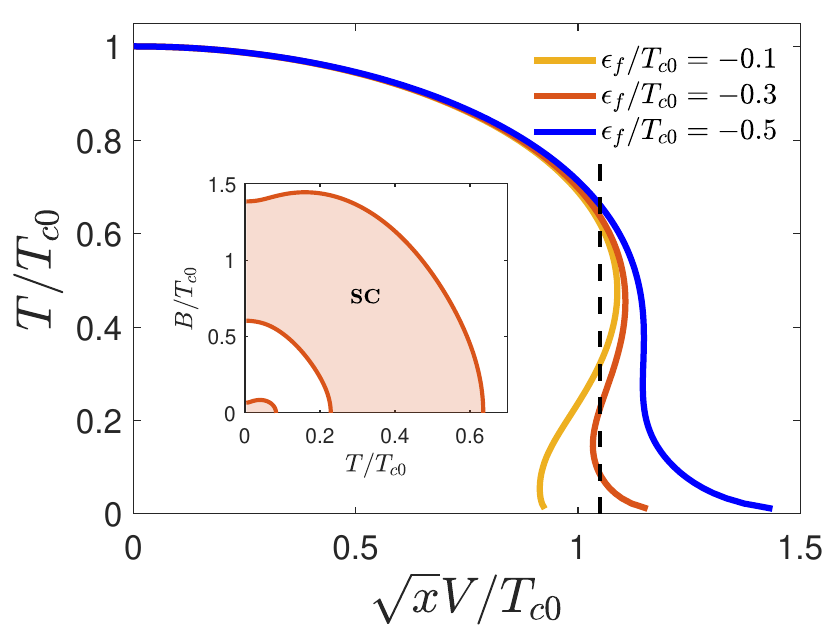}
\caption{Singlet superconducting transition temperature as a function of effective Kondo hybridization $\sqrt{x}V$ and impurity energy $\e_f$ in the absence of magnetic field and ASOC. The inset shows the calculated $B-T$ superconducting phase diagram with $\e_f/\Tco=-0.3$ and $\sqrt{x}V/\Tco=1.05$ (corresponding to the orange line at the intersection with the black dashed line).
}
\label{fig:ef_3Tc}
\end{figure}
%%%%%%%%%%%%%%%%%%%%%%%%%%%%%%%%%%%%%%

%%%%%%%%%%%%%%%%%%%%%%%%%%%%
\begin{figure*}[t]
\centering\includegraphics[width=7in]{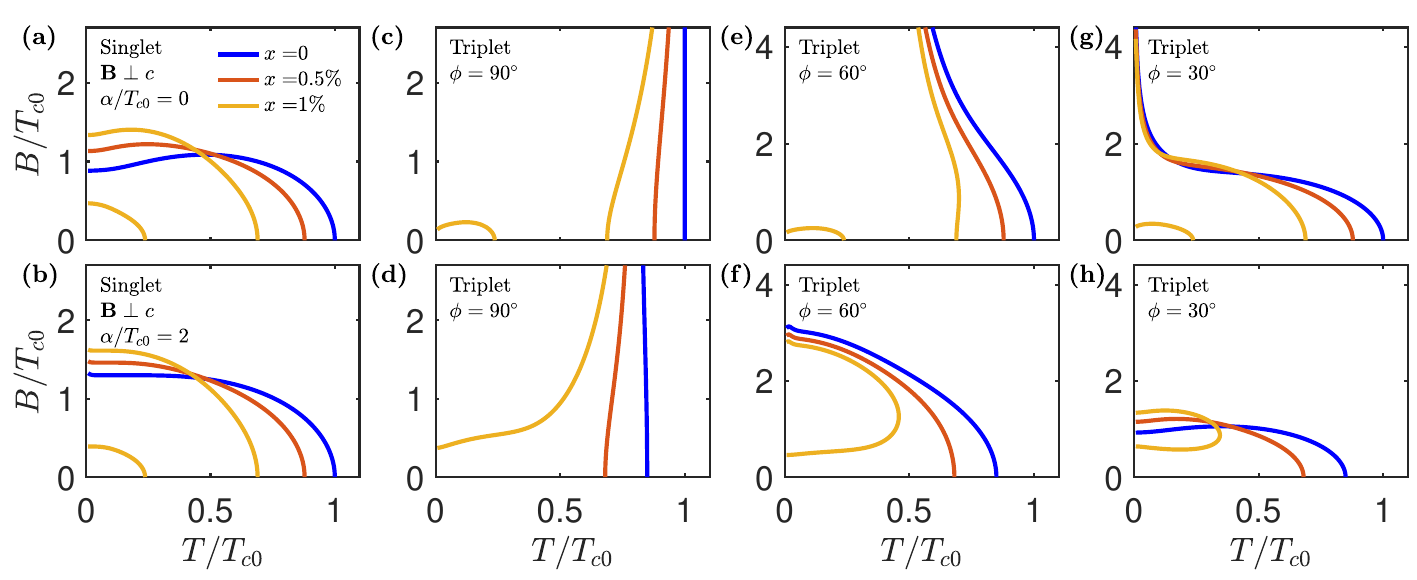}
\caption{Evolution of superconducting boundary with magnetic field direction, impurity concentration, and strength of spin-orbit coupling. The calculation utilizes $\e_f/\Tco=-0.1$ and $V/\Tco=10$. ASOC is chosen to break $c$-axis mirror, $\bgk=\alpha(-k_y,k_x,0)$ with the strength $\alpha/\Tco=0$ (the upper panel) and $\alpha/\Tco=2$ (the bottom panel).  (a)-(b) Magnetic field is spatial isotropic in $ab$ plane for singlet pairing. (c)-(h) Influence of magnetic field direction for triplet pairing with $\bdk=(k_y,0,0)$. The magnetic field is fixed in $ab$ plane with angle $\phi$ to $a$-axis.
}
\label{fig:Bdire_SOC_x}
\end{figure*}
%%%%%%%%%%%%%%%%%%%%%%%%%%%%%%%%%%%%%%

\section{Discussion of experimental data}\label{sec:exp}

\subsection{Singlet Superconductivity}

To show the validity of our superconducting model, we first compare the theoretical $B-T$ phase diagrams with experimental data. We note that we just intend to produce the observed $B_{c2}$ qualitatively rather than a quantitative fit to experimental data.  A more quantitative approach would require that we relax our assumption of a spherical Fermi surface in favor of a more materials-specific band structure.  This is not the goal of the present work, but rather to showcase the diversity of phase diagrams that emerge from our (simplified) theoretical model. 

We start with cobalt doped NbSe$_2$, where Co atoms are introduced by diffusion near the interface and provide localized $d$-electrons~\cite{Tingyu2023CoNbSe2}. Previous work suggests that monolayer NbSe$_2$ supports a mixture of singlet and triplet order parameters due to Ising spin orbit coupling~\cite{he2018nodalNbSe2,mockli2018NbSe2,sticlet2019FMchain_NbSe2} (that quickly decreases with increasing layers recovering in-plane inversion symmetry~\cite{KinFaiMak2016ising}).  It is therefore appropriate assume singlet pairing without ASOC in the theory to compare with the $\sim \! 10$ layer Co-NbSe$_2$ experimental data~\cite{Tingyu2023CoNbSe2}.  This is shown in the left panel of Fig.~\ref{fig:theo_exp}, where our theory is shown in panel (a) and the experimental data in panel (d).  Both theory and experiment show a window of superconductivity at intermediate magnetic fields and temperature.  At low magnetic field and low temperature we find a Kondo phase.  Increasing the magnetic field gives reenterant superconductivity -- a surprising result for s-wave superconductivity. 

We make some further observations: (i) With magnetic impurities, it is possible to have three transition temperatures ($T_{c1}>T_{c2} \: (\TK)>T_{c3}$); (ii) $T_{c1}$ is suppressed by magnetic doping which is a well-known conclusion~\cite{gorkov1961contributionEng}; (iii) Without ASOC the superconducting transition is isotropic in $\bB$; and (iv) In the zero temperature limit and without ASOC, the critical magnetic field can be simplified to 
\begin{equation}\label{Eq:Bc2}
    \ln{\frac{B}{\Tco}}+\frac{xV^2}{\e_f^2+xV^2}\frac{1}{2} \ln{\frac{|B^2-\e_f^2-xV^2|}{B^2}}-\ln{\frac{\pi}{2 e^{\gamma_0}}}=0,
\end{equation}
where $\gamma_0$ is Euler's constant.

Figure \ref{fig:ef_3Tc} shows the transition temperatures of singlet superconductors as a function of magnetic doping. Initially, $\Tc$ decreases with increasing magnetic doping. However, at higher doping,  particularly when the impurity energy $|\e_f|$ is small, an S-shaped $\Tc$ curve emerges. For some impurity concentrations multiple phase transitions are predicted (see the dashed line intersecting the orange line in Fig.~\ref{fig:ef_3Tc}).  The intermediate temperature phase set by  
$T_{c2}\sim$ $\TK$ yields a resistive state attributed to the Kondo interaction that enhances pair-braking and disrupts superconductivity. As the temperature is further lowered, a third critical temperature $T_{c3}$ for reentering superconductivity is predicted when the pair-breaking passes through its maximum and then diminishes, allowing superconductivity to reemerge.

The corresponding $B-T$ phase diagram shows three distinct phases (see the inset of Fig.~\ref{fig:ef_3Tc}). 
A Kondo-induced resistive phase is nestled in between two superconducting regions. In the zero temperature limit, the three critical fields given by Eq.~\eqref{Eq:Bc2} satisfy 
$B_3^c<|\e_f|<B_2^c<\sqrt{\e_f^2+xV^2}<B_1^c$.  A maximum of pair-breaking strength is observed at $B=|\e_f|$, corresponding to the energy required to excite local electrons to the Fermi level. However, when $|\e_f|$ becomes large, the pair-breaking effect is insufficient to disrupt superconductivity, resulting in a single critical temperature.

We note that in the paramagnetic limit (i.e. the highest critical field $B_{1}^c$ that breaks superconductivity), the superconducting state is destroyed due to the alignment of electron spins.  We find that the paramagnetic limit increases with doping, and exceeds $\sqrt{\e_f^2+xV^2}$ (see Fig.~\ref{fig:Bdire_SOC_x}(a)). As doping increases, $\Tc$ decreases while $B_1^c$ rises, and a Kondo-induced resistive states appears. This trend is consistent with experimental observations~\cite{Tingyu2023CoNbSe2}. 

\subsection{Triplet Superconductivity}

Before we can apply our framework to triplet superconductors, we first need to determine the appropriate order parameter $\bd$ and spin-orbit coupling $\bgk$.  For UTe$_2$, the pairing symmetry remains controversial despite several attempts to determine it experimentally using specific-heat~\cite{kittaka2020orientation}, scanning tunneling microscopy~\cite{jiao2020chiralkondoUTe2}, Kerr effect~\cite{paglion2021science}, Knight shift~\cite{aoki2022Knight,aoki2023knight}, and pulse echo ultrasound~\cite{theuss2024single}.  For this work, we adopt a more phenomenological approach.  For UTe$_2$ we assume the form $\bd=(k_y+ik_z,k_x,ik_x)$ and note that this choice is compatible both with some of the experiments on UTe$_2$ (e.g. Refs.~\cite{jiao2020chiralkondoUTe2,lee2023ute2}), and also with theoretical considerations starting from the D$_{2h}$ point group symmetry (e.g. Refs.~\cite{roising2024thermodynamic,sigrist1991phenomenological}).

With this choice of $\bd$, we get the lowest upper critical field in the $a$ direction, with a large $d_x$ component. It has a net magnetization $i\braket{\bd\times\bd^*}_\bk$ along the $a$-axis (the magnetic easy axis of UTe$_2$~\cite{aoki2019reviewUbased}). Generically, $\bd$ does not need to be fixed and could depend on the magnetic field.  For example, some recent works~\cite{aoki2021anisotropic,aoki2021inhomogeneous,machida2021nonunitary} consider the rotation of $\bd$ in UTe$_2$ with an applied magnetic field.  This possibility is not ruled out by our results, and for $B \gtrsim 20~T$, we expect some alignment of $\bd$ with the applied field.  However, in this work we focus on the low field regime where the direction of $\bd$ is determined only by the easy axis of the UTe$_2$ crystal.  As we observe below, the qualitative agreement between our theory and experimental data indicates that $\bd$ vector rotation is not necessary to understand the phase diagram.

It is known that UTe$_2$ shows a coexistance of Kondo resonance and superconductivity where the Kondo temperature varies from $\sim$19.6 K to 26 K~\cite{jiao2020chiralkondoUTe2} and $\Tc=1.6\sim2.1$ K depending on sample preparation~\cite{ran2019nearly_UTe2,aoki2019unconventionalUTe2,aoki2023knight}. The U-$5f$ electrons carry the magnetic moment and therefore $x\equiv1$. The antisymmetric spin-orbit coupling arising from the local inversion symmetry breaking at uranium
atoms~\cite{ishizuka2021PAMUTe2} is simplified to $\bgk=\alpha(-k_y,k_x,0)$. The finite strength of ASOC also helps to suppress the triplet superconductivity since it is not parallel to the $\bd$ vector. 

Fig.~\ref{fig:theo_exp}(b) shows the calculated $B_{c2}$ with $\bB\parallel a$ and $\bB\parallel b$ using these assumptions. We use a lower impurity energy $\epsilon_f$ to make $\TK>\Tc$. For $\bB\parallel a$, SC is quickly destroyed due to the large $a$-component of $\bd$ vector. However, we get a $L$-shaped $B_{c2}$ curve for $\bB\parallel b$, where the inflection point happens at $B\approx|\epsilon_f|$ with the strongest Kondo-induced pair-breaking strength. A higher magnetic field overcomes the corresponding excitation energy and screens out the Kondo effect, resulting in the increase of $\Tc$. We note that in our model the $B_{c2}$ curve along $c$-axis is slightly higher than that along $b$-axis, while it is lower than $b$-axis $B_{c2}$ in the experiment.  We attribute this discrepancy to our simplifying assumption of a spherical Fermi surface which does not take into account the large deformation of the density of states caused by the large $c$ lattice constant (see e.g. Ref.~\cite{ishizuka2021PAMUTe2}).  Nonetheless, the qualitative agreement is still very good.

Finally, we look at URhGe.  Here we assume $\bd=(k_z,ik_z,k_x+ik_y)$.  This choice has a net magnetization along the $c$-axis and is consistent with previous theoretical~\cite{mineev2002sc} and experimental~\cite{huxley2004urhge} work on this material. URhGe exhibits a ferromagnetic (FM) order aligned to the $c$-axis at Curie temperature $T_{Curie}=9.5$ K and enters SC phase at $\Tc=0.25$ K with FM state persisting~\cite{aoki2019reviewUbased}. The field reentrant superconductivity is suggested to be strongly linked to the Kondo interation~\cite{suzuki2020kondo}. The ASOC of URhGe arises from breaking local inversion symmetry~\cite{yanase2014zigzag,tada2016SOCdvecUCoGe} and is taken to be $\bgk=\alpha(0,k_x,0)$.  Our results are shown in  Fig.~\ref{fig:theo_exp}(c) computed for $\bB\parallel b$-axis, and  compared to the measurement Fig.~\ref{fig:theo_exp}(f).  We choose a larger hybridization energy than UTe$_2$ since the bond coupling U-5$f$ and Rh-4$p$ electrons in URhGe~\cite{divivs2002U5fRh4p} is shorter than for the U-5$f$ and Te-5$p$ electrons in UTe$_2$~\cite{kang2022U5fTe5p}.  With this strong hybridization, $\Tc$ at zero field is significantly suppressed by the Kondo effect. However, this suppression is then mitigated at high fields, leading to a $\Tc$ that exceeds its zero-field value. We also see two separate superconducting regions in theory. This is due to antisymmetric spin-orbit coupling that partly enhances the pair-breaking strength and breaks superconductivity around $B=|\epsilon_f|$. At even higher fields, the Kondo-induced pair-breaking is weakened, leading to the high-field superconducting phase.

\section{Additional predictions of the model}

We can now explore other possible phase diagrams by varying the parameters in the theory with the hope that these might be observed experimentally in other materials.  Assuming that ASOC breaks the $c$-axis mirror symmetry, we can generically write $\bgk=\alpha(-k_y,k_x,0)$~\cite{gor_rashba2001superconducting,sigrist2004superconductivity}. For singlet superconductors, we assume a constant superconducting gap ($\hat{\psi}_\bk=1$) 
similar to the results shown in Figs.~\ref{fig:Bdire_SOC_x}(a) and (b). (We note that the previous results without ASOC holds for any $\hat{\psi}_\bk$. However, the k-dependent $\bgk$ will be different depending on the singlet order parameter.) The superconductivity remains isotropic with respect to the field direction in the $ab$ plane. $\Tc$ at zero field is completely unaffected by ASOC since $\boldsymbol{\gamma}_{-\bk}=-\bgk$, and the effective field vector $\bb$ vanishes in Eq.~\eqref{Eq:main}. The upper critical field $B_{c2}(0)$ increases due to the presence of $\bgk\not\parallel \bB$, which weakens pair-breaking, thereby requiring a higher magnetic field to destroy superconductivity.

For triplet superconductors, we consider a triplet order parameter $\bdk=(k_y,0,0)$ with $\braket{S_x}=0$ for simplicity. We focus on the $ab$ plane magnetic field with angle $\phi$ to the $a$-axis. 
Figures.~\ref{fig:Bdire_SOC_x}(c)-(h) show the $B-T$ phase diagrams of triplet superconductors as a function of magnetic doping and antisymmetric spin-orbit coupling. $\Tc$ at zero field is suppressed by doping as expected. However, surprisingly, 
$\Tc$ is enhanced by the magnetic field when $\bB\perp\bdk$ since the magnetic field weakens the Kondo-induced pair-breaking through the $|\hat{d}_\bk|^2$ term in Eq.~\eqref{Eq:main}, screening out the magnetic impurities (see Fig.~\ref{fig:Bdire_SOC_x}(c) and (d)).  To our knowledge, this mechanism for the enhancement of $\Tc$ has not been reported previously in the literature.

Rotating the $\bB$ direction to the $\bd$ direction results in a suppression of the $\Tc$ curve (see Figs.~\ref{fig:Bdire_SOC_x}(e)-(h)). Provided $\bB$ is nearly parallel to $\bd$, the increase of $\hat{b}\cdot \bdk^*$ term enhances the pair-breaking and reduces $\Tc$.  We can therefore predict the necessary conditions for the existence of $B_{c2}(0)$ at zero temperature: (i) If $\bgk\cdot \bB=0$ for all $\bk$, then it requires $\bdk\parallel \bB$ for all $\bk$ (see the zero ASOC case in Figs.~\ref{fig:Bdire_SOC_x}(c), (e) and (g), where only $\phi=0$ gives $B_{c2}(0)$); (ii) If $\bgk\cdot \bB\neq 0$ for any $\bk$, then $B_{c2}(0)$ exists provided $\bB\cdot\bdk\neq0$ for any $\bk$ (see Figs.~\ref{fig:Bdire_SOC_x}(d), (f) and (h)). This is because ASOC enhances the $\hat{b}\cdot \bdk^*$ term which in turn increases the pair-breaking strength, yielding a zero-temperature upper critical field.  We hope that these conditions will help experimentalists correctly identify the pairing symmetry of new and existing triplet superconductors.  In sharp contrast to singlet superconductors, the presence of ASOC reduces the zero field triplet $\Tc$, since pair-breaking in the $|\hat{d}_\bk|^2$ term in Eq.~\eqref{Eq:main} (triplet) is increased.   

We predict that higher doping concentrations will make it easier to observe the Kondo resistive state at low temperature in triplet superconductors. In the absence of ASOC, the Kondo phase appears at low magnetic fields, while superconductivity reenters at higher fields.  However when ASOC and Kondo hybridization are both strong, pair-breaking is significantly enhanced, completely suppressing superconductivity at zero field (see Figs.~\ref{fig:Bdire_SOC_x}(d), (f) and (h)). Applying a magnetic field then screens out the magnetic impurities, resulting in a field-induced superconductivity.  Enhancing superconductivity with a magnetic field is very unusual. This suggests that materials not typically considered superconducting, but which exhibit a strong Kondo effect and spin-orbit coupling, could potentially be made superconducting through the application of a magnetic field.

Finally, the last term of Eq.~\eqref{Eq:main} (triplet) arises from the non-unitary $\bd$ vector, which generates a net spin triplet polarization $\bS_t(\bk)=i\braket{\bdk\times\bdk^*}$. This term vanishes in the absence of Kondo hybridization. With Kondo hybridization present, the pair-breaking from this term is governed by the interaction energy $\bS_t \cdot \bb$. It stabilizes superconductivity when $\braket{\bS_t}$ is (more) anti-parallel to the magnetic field. 

\section{Conclusion}\label{sec:conclu}
Inspired by the intriguing $B-T$ phase diagram in systems like engineered Co-NbSe$_2$ and U-based triplet superconductors, we extend the periodic Anderson model to the case of spin triplet superconductors including both magnetic fields and spin-orbit coupling.  Using the Green's function formulation, we obtain the linearized gap equation for the superconducting transition.  We find a Kondo resistive state existing below the superconducting phase for both singlet and triplet superconductors that is characterized by small impurity energy $|\e_f|$. Conversely, in cases of large $|\e_f|$, the Kondo temperature is larger than the superconducting $\Tc$, resulting in a single critical temperature. While singlet pair-breaking is isotropic with the applied magnetic field direction, triplet superconductivity is strongly field anisotropic, remaining superconducting unless $\bB\parallel\bdk$ for all $\bk$. However, including an antisymmetric spin-orbit coupling imposes a paramagnetic limit to triplet superconductivity provided $\bB\cdot\bdk\neq 0$ for any $\bk$.  Our study illustrates how the Kondo interaction stabilizes the non-unitary triplet order parameter, that is intricately linked to both $\bB$ and $\bgk$.  Beyond exploring hypothetical combinations of order parameters, magnetic impurities and spin-orbit interaction, we also directly apply our framework to two known Kondo superconductors including UTe$_2$ and URhGe, as well as the engineered Co-NbSe$_2$.  Our theoretical framework can provide good qualitative agreement with the observed phase diagrams in these different materials.  This highlights the pivotal role played by the Kondo interaction in such magnetic superconductors.

%\newpage
\begin{acknowledgments}
We thank Sheng Ran for helpful comments. This work was supported by the Singapore National Research Foundation Investigator Award (NRF-NRFI06-2020-0003).  B\"O acknowledges support from the Singapore NRF Investigatorship (Grant No. NRF-NRFI2018-8), Competitive Research Programme (Grant No. NRF-CRP22-2019-8), and MOE-AcRF-Tier 2 (Grant No. MOE-T2EP50220-0017).

\end{acknowledgments}
%\newpage
%%%%%%%%%%%%%%%%%%%%% appendix A %%%%%%%%%%%%%%%%%%%%%%%%%%%%%%%

\appendix
\section{Normal state Green's function}\label{sec:appendix_Green}

In this Appendix we provide a detailed derivation of the Green's functions. The Green's functions in the imaginary-time momentum space are defined as~\cite{sigrist1991phenomenological}
\begin{equation}\label{Eq:Green}
G_{\mathcal{AB}}^{s_1 s_2}(\bk,\tau)=-\braket{T_\tau \mathcal{A}_{\bk s_1}(\tau)\mathcal{B}_{\bk s_2}^\dag(0)},
\end{equation}
where $\mathcal{A}$ and $\mathcal{B}$ represent conduction electron $c$ or localized electron $f$ operators. $s_1$ and $s_2$ are spin indices. After transforming to Matsubara frequencies, Green's function can be obtained by solving the Dyson equation $(g^{-1}-\mathcal{V})G=I$, which is
\begin{equation}\label{Eq:Dyson}
\begin{split}
\left(\begin{array}{cc}
\iwn-\e_f-\bB\cdot \boldsymbol{\sigma} & -\sqrt{x}V \\
-\sqrt{x}V & \iwn-\e_\bk-(\bgk+\bB)\cdot \boldsymbol{\sigma}
\end{array}\right)\\
\times\left(\begin{array}{cc}
G_{ff} & G_{cf} \\
G_{fc} & G_{cc}
\end{array}\right)=I_{4\times4}.
\end{split}
\end{equation}
The Green's function for conduction electrons is obtained as
\begin{equation}\label{Eq:Green_solu}
G_{cc}(\bk,\iwn)=G_+(\bk,\iwn)+\hat{b}\cdot\boldsymbol{\sigma}G_-(\bk,\iwn),
\end{equation}
with $G_+$ and $G_-$ given by
\begin{eqnarray}
    G_+(\bk,\iwn)=&&\frac{\iwn-\e_\bk+xV^2\frac{(\e_f-\iwn)}{(\e_f-\iwn)^2-B^2}}{(\e_\bk-\xi_\bk^{+})(\e_\bk-\xi_\bk^{-})},\\
    G_-(\bk,\iwn)=&&\frac{b}{(\e_\bk-\xi_\bk^{+})(\e_\bk-\xi_\bk^{-})}.
\end{eqnarray}
The poles of the Green's function are written as
\begin{equation}\label{Eq:pole}
    \xi_\bk^{\pm}=\iwn+xV^2\frac{(\e_f-\iwn)}{(\e_f-\iwn)^2-B^2} \pm b.
\end{equation}
The Fermi surface splits into two different sheets with the presence of a magnetic field, hybridization, and ASOC.

\section{Linearized gap equation}\label{sec:appendix_gapeq}
In this section, we solve the linearized gap equation. Eq.~\eqref{Eq:Gapeq} gives rise to 
\begin{eqnarray}\label{Eq:gapeq_psi}
    \psi_\bp=&&-T\sum_{n\bk}V_{\bp\bk}\bigg[ \Big(G_+G_+-G_-G_-\hat{b}\cdot\hat{\tb}\Big)\psi_\bk \non\\
    &&+\Big(G_-G_+\hat{b}-G_+G_-\hat{\tb}+iG_-G_-(\hat{b}\times\hat{\tb})\Big)\cdot\bdk \bigg],\\ \label{Eq:gapeq_d}
    \bdk=&&-T\sum_{n\bk}V_{\bp\bk}\bigg[ \Big(G_-G_+\hat{b}-G_+G_-\hat{\tb}-iG_-G_-(\hat{b}\times\hat{\tb}) \Big)\psi_\bk \non\\
    &&+\Big(G_+G_++G_-G_-\hat{b}\cdot\hat{\tb} \Big)\bdk \non\\
    &&+i\Big(G_-G_+\hat{b}+G_+G_-\hat{\tb} \Big)\times\bdk \non\\
    &&-G_-G_-\Big((\bdk\cdot\hat{\tb})\hat{b}+(\bdk\cdot\hat{b})\hat{\tb} \Big)   \bigg],
\end{eqnarray}
Here we use the notation that $G_{\delta}G_{\eta}=G_{\delta}(\bk,\iwn)G_{\eta}(-\bk,-\iwn)$ with $\delta,\eta=\pm$. The singlet and triplet gap equations are weakly coupled to each other. However, this coupling depends on the degree of particle-hole asymmetry and is of the order $|\bgk|/\e_F<<1$ ($\e_F$ is Fermi energy)~\cite{sigrist2004superconductivity}. This coupling becomes exactly zero under our spherical Fermi surface assumption. We therefore ignore this coupling and solve singlet and triplet gap equations separately. The singlet and triplet gap functions are written as~\cite{sigrist1991phenomenological}
\begin{eqnarray}\label{Eq:psi_expand}
    \psi_\bk=&&\sum_{m} c_m Y_{lm}(\Omega_\bk),\quad l\in even, \non\\ \label{Eq:d_expand}
    \bdk=&&\sum_{m,\hat{n}} c_{m,\hat{n}}Y_{lm}(\Omega_\bk)\hat{n},\quad l\in odd.
\end{eqnarray}
Using the potential $V_{\bp\bk}=-\sum_{l=0}^{\infty}4\pi V_l\sum_m{Y_{lm}(\Omega_\bp)Y_{lm}^*(\Omega_\bk)}$~\cite{anderson1961generalized,balian1963superconductivity}, Eqs.~\eqref{Eq:gapeq_psi} and~\eqref{Eq:gapeq_d} are simplified to
\begin{eqnarray}\label{Eq:psi_Tc}
    \frac{1}{V_{l\in even}}=&&\Tc\sum_{n\bk}\Big[G_+G_+-G_-G_-\hat{b}\hat{\tb} \Big]|\hat{\psi}_\bk|^2, \\  \label{Eq:d_Tc}
    \frac{1}{V_{l\in odd}}=&&\Tc\sum_{n\bk}\bigg[ \Big(G_+G_++G_-G_-\hat{b}\cdot\hat{\tb} \Big)|\hat{\mathbf{d}}_\bk|^2 \non\\ 
    &&+i\Big(G_-G_+\hat{b}+G_+G_-\hat{\tb} \Big)\cdot (\hat{\mathbf{d}}_\bk\times\hat{\mathbf{d}}_\bk^*) \non\\
    &&-2G_-G_-(\hat{\mathbf{d}}_\bk\cdot\hat{\tb})(\hat{\mathbf{d}}_\bk^*\cdot\hat{b})  \bigg].
\end{eqnarray}
We then calculate the superconducting transition temperature $T_c$ for singlet pairing and triplet pairing by evaluating the Cooper diagram~\cite{DeGennes2018superconductivity,gorkov2005kondo}. We introduce the bare superconducting critical temperature $\Tco$ without magnetic field, $f$-electron or ASOC. The gap equations for $\Tco$ are
\begin{eqnarray}\label{Eq:psi_Tc0}
    \frac{1}{V_{l\in even}}=&&\Tco\sum_{n\bk}g_{cc}(\bk,\iwn)g_{cc}(-\bk,-\iwn)|\hat{\psi}_\bk|^2, \\  \label{Eq:d_Tc0}
    \frac{1}{V_{l\in odd}}=&&\Tco\sum_{n\bk}g_{cc}(\bk,\iwn)g_{cc}(-\bk,-\iwn)|\hat{\mathbf{d}}_\bk|^2,
\end{eqnarray}
where $g_{cc}^{-1}(\bk,\iwn)=\iwn-\e_\bk$ is the bare Green's function. By combining Eqs.~\eqref{Eq:psi_Tc} and~\eqref{Eq:psi_Tc0} (Eqs.~\eqref{Eq:d_Tc} and~\eqref{Eq:d_Tc0}) for singlet (triplet), and taking into account the frequency cutoff $\omega_D$, we finally obtain the pair-breaking equation for $\Tc$, which is given in Eq.~\eqref{Eq:main}.

\bibliography{refPAMKondo}

\end{document}